\pdfoutput=1

\documentclass[11pt,twoside,a4paper,cmspaper,final,collab]{cms-tdr}
\def\svnVersion{144861}\def\cmsCernNo{2012-144}\def\cmsCernDate{\today}\def\cmsMessage{Submitted to Physics Letters B}

\begin{document}\cmsNoteHeader{HIN-11-008}

\hyphenation{had-ron-i-za-tion}
\hyphenation{cal-or-i-me-ter}
\hyphenation{de-vices}

\RCS$Revision: 144782 $
\RCS$HeadURL: svn+ssh://alverson@svn.cern.ch/reps/tdr2/papers/HIN-11-008/trunk/HIN-11-008.tex $
\RCS$Id: HIN-11-008.tex 144782 2012-08-24 12:01:59Z delacruz $
\providecommand{\sNN}{\ensuremath{\sqrt{s_{_\mathrm{NN}}}}\xspace}
\providecommand{\mt}{\ensuremath{m_\mathrm{T}}\xspace}
\providecommand{\HYDJET}{\textsc{hydjet}\xspace}
\providecommand{\PYTHHYD}{\PYTHIA{}+\HYDJET}
\newlength\cmsThreeFigWidth
\ifthenelse{\boolean{cms@external}}{\setlength\cmsThreeFigWidth{0.95\columnwidth}}{\setlength\cmsThreeFigWidth{0.58\textwidth}}
\newlength\cmsFigWidth
\ifthenelse{\boolean{cms@external}}{\setlength\cmsFigWidth{0.95\columnwidth}}{\setlength\cmsFigWidth{0.9\textwidth}}
\cmsNoteHeader{HIN-11-008} % This is over-written in the CMS environment: useful as preprint no. for export versions
\title{Study of \PW\ boson production in PbPb and pp collisions at $\sNN = 2.76\TeV$}

\date{\today}

\abstract{
A measurement is presented of \PW-boson production in PbPb collisions carried out at a nucleon-nucleon (NN) centre-of-mass energy $\sNN$ of $2.76$\TeV at the LHC using the CMS detector.
In data corresponding to an integrated luminosity of  7.3\mubinv,
the number of $\PW \rightarrow \Pgm \Pgngm$ decays is extracted in the region of muon pseudorapidity $\abs{\eta^\mu}<2.1$ and transverse momentum $\pt^\Pgm>25\GeVc$.
Yields of muons found per unit of pseudorapidity correspond to $(159 \pm 10 (\text{stat.}) \pm 12 (\text{ syst.})) \times 10^{-8}$ \PWp\ and $(154 \pm 10 (\text{stat.}) \pm 12 (\text{syst.})) \times 10^{-8}$ \PWm\ bosons per minimum-bias PbPb collision.
The dependence of \PW\ production on the centrality of PbPb collisions is consistent with a scaling of the yield by the number of incoherent NN collisions.
The yield of \PW\ bosons is also studied in a sample of \Pp\Pp\ interactions at $\sqrt{s}= 2.76\TeV$ corresponding to an integrated luminosity of 231\nbinv.
The individual \PWp\ and \PWm\ yields in PbPb and \Pp\Pp\ collisions are found to agree, once the neutron and proton content in Pb nuclei is taken into account.  Likewise, the
difference observed in the dependence of the positive and negative muon production on pseudorapidity is
consistent with next-to-leading-order perturbative QCD calculations.
}

\hypersetup{%
pdfauthor={CMS Collaboration},%
pdftitle={Study of W boson production in PbPb and pp collisions at sqrt(s[NN]) = 2.76 TeV},%
pdfsubject={CMS},%
pdfkeywords={CMS, physics, heavy-ions, W bosons}}

\maketitle %maketitle comes after all the front information has been supplied

\section{Introduction}

The hot and dense matter produced in heavy-ion (AA) collisions can be studied in a variety of ways. One approach is to compare AA to proton-proton (\Pp\Pp) collisions as well as to collisions of protons or deuterons with nuclei. Another way is to compare yields of particles whose properties are modified by the produced medium to those of unmodified reference particles in the same AA collisions.
Direct photons play the reference role at the Relativistic Heavy Ion Collider (RHIC)~\cite{Adler:2005ig} and, more recently, also at the Large Hadron Collider (LHC)~\cite{Chatrchyan:2011ip,CMS_gamma_jet}. However, their measurement is complicated by copious background from $\pi^0$ and $\eta$ meson decays, and by the presence of photons produced in fragmentation processes of final-state partons that can be affected by the medium~\cite{Arleo:2006xb}. At LHC energies, new and cleaner references such as weak bosons in their leptonic decay modes become available~\cite{Kartvelishvili:1995fr,Dainese:PLB,ConesadelValle:2009vp}. The ATLAS and CMS collaborations recently reported first observations of \cPZ\ bosons in heavy-ion interactions, showing that their yields per nucleon--nucleon (NN) collision are essentially unmodified by the medium~\cite{Atlas:2010px,Chatrchyan:2011ua}.

Weak-boson production is recognised as an important benchmark process at hadron colliders.
Measurements at 7\TeV centre-of-mass (CM) energy in \Pp\Pp\ collisions at the LHC~\cite{CMS:2010xn,Chatrchyan:2011ig,Chatrchyan:2011jz,Chatrchyan:2011ne,Aad:2010yt,Aad:2010pg,Aad:2011dm,Aad:2011yn} and previously, at other hadron colliders (Tevatron~\cite{D0_Wprod,CDF_Wprod}, RHIC~\cite{STAR_Wprod,Phenix_Wprod} and S{\Pp\Pap}S~\cite{UA1_Wprod,UA2_Wprod}) with various collision energies,
 are well described by calculations based on higher-order perturbative quantum chromodynamics (QCD) using recent parton distribution functions (PDF). In PbPb collisions, \PW-boson production can be affected by initial-state conditions~\cite{Kartvelishvili:1995fr,Vogt:2000hp,Zhang:2002yz,Paukkunen:2010qg}, such as
 the mix of protons and neutrons. Since the leading-order \PW-production processes $\cPqu\cPaqd \rightarrow \PWp$ and
$\cPqd \cPaqu \rightarrow \PWm$ reflect mainly interactions that take place between valence quarks and sea antiquarks, the individual \PWp\ and \PWm\ rates are expected to be modified relative to \Pp\Pp\ collisions, but not their sum. This is often referred to as the isospin effect, as it stems from a different content of u and d quarks in the proton relative to lead nuclei. The PDF can also be modified in nuclei, as parton depletion (or shadowing) could change the yield of \PW\ bosons at the LHC by as much as 15\% in certain regions of kinematics~\cite{Paukkunen:2010qg}. Precise measurements of \PW\ production in heavy-ion collisions can therefore constrain the nuclear PDF and, moreover, provide insight into the PDF for neutrons.

The $\PW \rightarrow l \cPgn_l$ decays are of particular interest, since the charged leptons ($l$) lose negligible energy in the produced medium, regardless of its nature (partonic or hadronic) or specific properties~\cite{Dainese:PLB,ConesadelValle:2009vp}.
Since they are dominantly created from a left-handed valence quark and a right-handed sea antiquark, \PW\ bosons are mostly left-handed and emitted in the valence quark direction, thus towards non-zero rapidity. The \PWp\ decays to a left-handed neutrino and a right-handed positive lepton, which is thus boosted back towards midrapidity, while the \PWm\ decays to a left-handed negative lepton which is boosted towards higher rapidity.
This fact creates a difference in $l^+$ and $l^-$ yields as a function of lepton pseudorapidity, $\eta$, defined as $\eta = -\ln[\tan(\theta/2)]$, $\theta$ being the polar angle of a particle trajectory with respect to the direction of the anticlockwise-circulating heavy-ion beam.
This angular difference and the relative abundances of \PWp\ and \PWm\ bosons produced in PbPb compared to \Pp\Pp\ collisions (isospin effect) manifests itself in a lepton charge asymmetry, defined as a difference in $l^+$ and $l^-$ contributions divided by their sum.
The measurement of this asymmetry as a function of muon pseudorapidity is quite robust, as it is insensitive to many systematic uncertainties.      %Given the V-A nature of the electroweak interaction, the possible $\pm1$ helicity states of $W^+$ and $W^-$ and the fixed neutrino helicity, the W leptonic decay produces well determined lepton angular distributions, non-symmetric between both charges. The muon charge asymmetry accounts for this angular differences as well as the relative abundance of $W^+$ and $W^-$ bosons.
The \PW\ bosons appear therefore to be well suited to probe the characteristics of the initial state of PbPb collisions at LHC energies.

This Letter reports the observation of \PW-boson production in a minimum bias (MB) sample of N$_\mathrm{MB}=55.7 \times 10^6$ events from PbPb collisions collected with the Compact Muon Solenoid (CMS) detector at the CM energy for colliding nucleon pairs of $\sNN=2.76\TeV$. This sample corresponds to an integrated luminosity of $(7.3 \pm 0.3)\mubinv$. These data were recorded during the first PbPb LHC data taking period at the end of 2010. In addition, we present results of a comparison analysis of \PW\ production in \Pp\Pp\ interactions in data obtained at the same $\sNN$ for an integrated luminosity of $(231 \pm 14)\nbinv$, which is of a similar size to the nucleon-nucleon equivalent luminosity of the PbPb data.

The Letter is organised as follows: the CMS detector is briefly described in Section~2, followed by the description of the experimental methods used for online and offline data selection in the PbPb and \Pp\Pp\ collected samples of events. The Monte Carlo (MC) simulations and the acceptance and efficiency correction factors derived from them are described there as well. The results and their discussion, together with comparison with theoretical predictions, are presented in Section~3. Finally the conclusions of this study are summarised in Section~4.

\section{Experimental methods}

A detailed description of the CMS detector can be found in Ref.~\cite{CMS:2008zzk}. In brief, a silicon pixel and strip tracker is located within a superconducting solenoid of 6\unit{m} internal diameter that provides a magnetic field of 3.8\unit{T}. The tracker consists of 66 million pixel and 10 million strip-detector channels, used to measure charged-particle trajectories for $\abs{\eta} < 2.5$.
It provides a vertex resolution of ${\approx}15\mum$ in the transverse plane. Located within the solenoid, but outside of the tracker, are a crystal electromagnetic calorimeter and a brass/scintillator hadron calorimeter. Muons are measured within $\abs{\eta}< 2.4$ in gaseous detector planes embedded in the steel return yoke of the magnet. A matching of outer muon trajectories to the tracks measured in the silicon tracker provides a transverse momentum (\pt) resolution between 1 and 2\%, for \pt\ values up to 100\GeVc. In addition, CMS has extensive forward calorimetry, in particular, two steel/quartz-fibre Cherenkov, forward hadron calorimeters (HF), on each side of the collision point, covering $2.9 < \abs{\eta} < 5.2$.

The centrality of PbPb collisions reflects the geometric overlap (impact parameter) of the incoming nuclei, and is related to the energy released in these collisions and the effective number of NN interactions. CMS defines the centrality of a PbPb collision through bins that correspond to fractions of the total hadronic inelastic cross section, as observed in the distribution of the sum of the energy deposited in the HF~\cite{HIN-10-004,D'Enterria:2007xr}. The five bins in centrality used in this analysis, ordered from the smallest to the largest energy deposited in the HF, range from the most peripheral, 50--100\%, 30--50\%, 20--30\%, 10--20\%, to the most central, 0--10\%, collisions. These bins can be related through a Glauber model~\cite{Miller:2007ri} to the number of nucleon-nucleon collisions per event.

In this analysis, \PW\ bosons are measured through their $\PW \rightarrow \Pgm \Pgngm$ decays. Muons can be cleanly identified and reconstructed, despite the high-multiplicity environment of heavy-ion collisions, a fact that makes this channel particularly suitable for measuring \PW\ production.
The muon charge and transverse momentum vector are evaluated from the curvature of the track in the silicon tracker. The neutrino is not detected, but a large imbalance in the vector sum of the transverse momenta of all charged particles measured in the tracker is used to signal its presence.

A sample of MB events is selected that have a reconstructed primary vertex based on at least two tracks, and an offline-determined coincidence of energy depositions in both HF calorimeters, with at least three towers, each above 3\GeV. These criteria reduce contributions from single-beam interactions with the environment (\eg beam-gas and beam-halo collisions within the beam pipe), ultra-peripheral electromagnetic collisions and cosmic-ray muons. The acceptance of this selection corresponds to $(97 \pm 3)$\% of the hadronic PbPb inelastic cross section~\cite{HIN-10-004}.

Events for this analysis are selected using the two-level trigger of CMS. At the first (hardware) level, one muon candidate with a \pt\ of at least 3\GeVc is required in the muon detectors. At the software-based higher level, one reconstructed track with a more precisely determined $\pt> 3\GeVc$ is again required in the muon detectors. For muons from \PW-boson decays, the single-muon trigger efficiency is estimated as
$(97.0 \pm 2.3)$\%.

Muon offline reconstruction has ${\approx}99\%$ efficiency to find tracks when hits in the muon detectors are taken as seeds. These tracks (called stand-alone muons) are matched to tracks reconstructed in the silicon tracker by means of an algorithm optimised for the heavy-ion environment~\cite{Roland:2006kz,D'Enterria:2007xr}. For a muon from \PW\ decays, the silicon-tracking efficiency is ${\approx}85\%$, which is less than for \Pp\Pp\ collisions, as track reconstruction in the PbPb environment requires more pixel hits to reduce the number of possible combinations resulting from large particle multiplicities. Combined fits of the stand-alone muon and tracker trajectories (called global muons) are used in extracting the results of this analysis. Muon pseudorapidities are restricted to $\abs{\eta^\Pgm}<2.1$, which provides uniform and good resolution both at the trigger stage and in offline reconstruction.

\begin{figure}[hbtp]
  \begin{center}
  \includegraphics[width=\cmsThreeFigWidth]{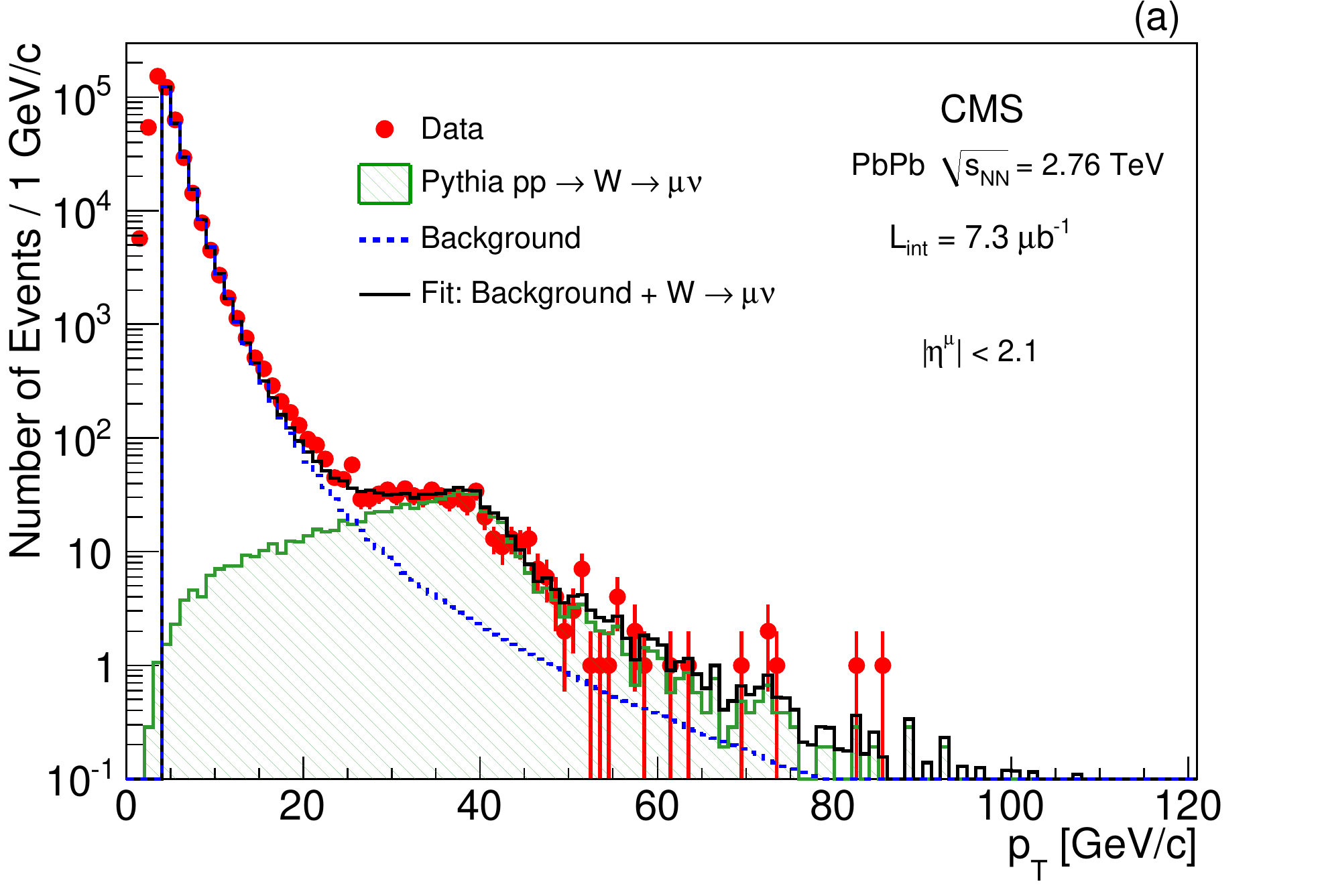} % 0.95 for PRL (was width=0.66)
    \includegraphics[width=\cmsThreeFigWidth]{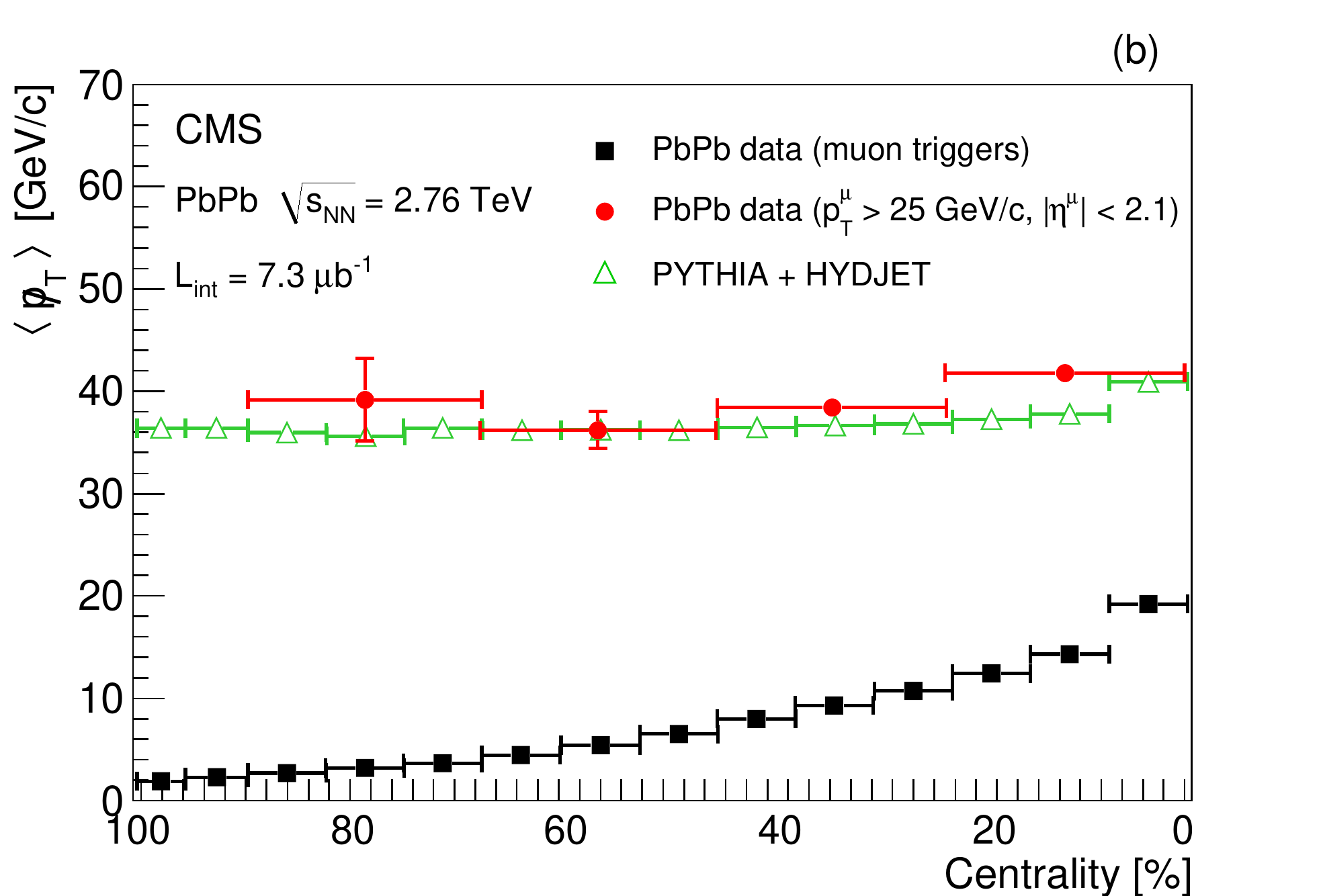} % 0.95 for PRL
    \includegraphics[width=\cmsThreeFigWidth]{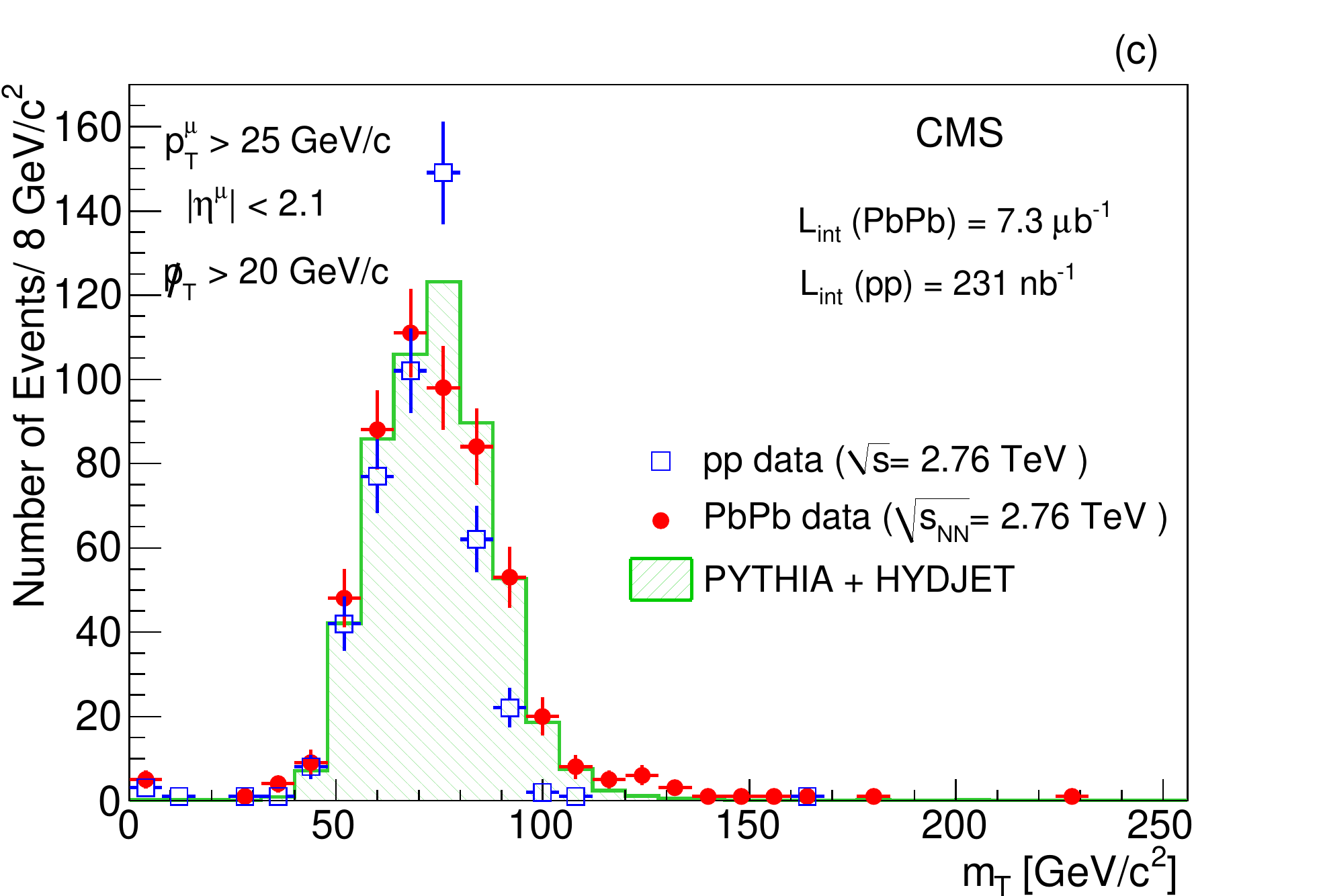} % 0.95 for PRL
    \caption{(a) Single-muon transverse-momentum spectrum for $\abs{\eta^\Pgm}<$2.1 in PbPb data (red-filled circles). Signal (green-hatched histogram) and background (blue-dashed histogram) contributions are fitted (black solid line) to the data. (b) Mean value of \PTm\ for charged tracks as a function of centrality, before any event selection is applied on the muon-triggered data (black squares) and after it (red-filled circles), together with predictions from the \PYTHIA{}+\HYDJET samples (green triangles). (c) Transverse mass distribution for selected events in PbPb (red-filled circles) and \Pp\Pp\ (blue open squares) data, compared to simulation (green-hatched histogram). The error bars represent statistical uncertainties. (See the text for more details).}
    \label{fig:pt}
  \end{center}
\end{figure}

A \cPZ-boson veto is applied to reject events that contain a second muon of opposite charge with $\pt > 10\GeVc$ that forms a dimuon invariant mass of $60< m_{\Pgm\Pgm} < 120\GeVcc$.
Background muons from cosmic rays and heavy-quark semileptonic decays are rejected by requiring a transverse impact parameter of less than 0.3~mm relative to the measured vertex. No muon isolation criteria are required. The single-muon \pt\ spectrum following this selection is shown in Fig.~\ref{fig:pt}(a) with red-filled circles.
The enhancement in the number of muons with $\pt > 25\GeVc$, expected from the decay of \PW\ bosons (green-hatched histogram), is evident. Details on the fit to the data are given below.

To further characterise events with muons arising from \PW\ decays, the imbalance ($\PTm$) in the sum of the charged-particle transverse momenta with $\pt>3\GeVc$ is computed for each event. The mean value of this transverse-momentum imbalance as a function of centrality of the PbPb collision is presented in Fig.~\ref{fig:pt}(b) for data (black-filled squares) selected with the two-level muon trigger described above. The presence of significant \PTm\ in central events
is expected as these events contain many particles that are not included in the sum, such as neutrals or charged particles produced at low transverse momentum or at large pseudorapidity.
For peripheral collisions, the net \PTm\ tends to be quite small. Once a high-$\pt$ muon is required in the data (red-filled circles), the $\langle \PTm \rangle$ shifts to higher values of ${\approx}40\GeVc$, and is far less dependent on the centrality of the collision. This agrees with expectations (green triangles) for \PTm\ values of undetected neutrinos originating from \PW\ decay. To enhance the contribution from the \PW\ signal, events are therefore required to have $\pt^\Pgm > 25\GeVc$ and $\PTm > 20\GeVc$.

The distribution in transverse mass (\mt) for the $\Pgm$ and \PTm\ system, computed as
\linebreak[4]$\mt = \sqrt{2 \pt^\Pgm \PTm ( 1 - \cos \phi)}$, where $\phi$ is the
azimuth between the $\pt^\Pgm$ and the \PTm\ vectors, is presented in Fig.~\ref{fig:pt}(c) (red-filled circles), together with the expectation from a sample of simulated \PW\ events (green-hatched histogram)
generated with \PYTHIA v.6.4~\cite{Ref:Pythia} that are passed through a detailed simulation of the CMS detector based on \GEANTfour~\cite{Agostinelli:2002hh}. The effect of larger background from the underlying event in PbPb collisions is taken into account by embedding detector-level signals from the simulated \PWp\ and \PWm\ decays into PbPb events generated with \HYDJET~\cite{Lokhtin:2005px} (referred to as \PYTHHYD samples in the rest of the Letter).
The \mt\ spectrum of Fig.~\ref{fig:pt}(c) is expected to have a sharp falling edge at the mass of the \PW\ boson, which is smeared by experimental resolution. The region of $\mt > 40\GeVcc$ is used to define the \PW\ signal.
Following this final selection,
a total of 275 $\Pgmp$ and 264 $\Pgmm$
events remain in the data sample.

 Residual contributions from \cPZ\ bosons with a misidentified muon or or a muon emitted in an insensitive region of the detector, as well as contributions from $\PW \rightarrow \Pgt \Pgngt$ processes, where the $\Pgt$ decays into $\Pgm \Pgngm \Pgngt$, are estimated using \Pp\Pp\ events simulated with \PYTHIA at the corresponding CM energy. A total estimated background contamination of 2.1\% of the selected sample, based on the \Pp\Pp\ MC simulation, is subtracted from the data,
as such electroweak processes are expected to scale with the number of elementary nucleon-nucleon collisions~\cite{Miller:2007ri}.
Remaining contaminations from a variety of QCD processes (mainly from semileptonic heavy-quark decays) in both \Pp\Pp\ and PbPb data, are
estimated by extrapolating the \mt\ distribution for both isolated and non-isolated, muon-enriched samples into the regions of signal.
The estimate from \Pp\Pp\ data provides an upper limit on possible contamination of the PbPb sample,
since parton energy loss (jet quenching)~\cite{HIN-10-004} in heavy-ion collisions can only lower the yields relative to \Pp\Pp\ production. As a cross-check, the same method is applied to PbPb data.
 In both cases, upper limits on the contamination from QCD processes of 1\% of the total selected PbPb sample are established, and no additional correction is applied for residual background, but a 1\% systematic uncertainty is attributed to these sources of background.

A cross-check on the contributions from signal and background to the selected sample is obtained by fitting the muon $\pt$ spectrum in Fig.~\ref{fig:pt}(a) to two components: one arising from $\PW \rightarrow \Pgm \Pgngm$ signal (a template is taken from \PYTHIA-simulated \Pp\Pp\ events), and another used for the background, modelled by a modified Rayleigh function~\cite{CMS:2010xn} with 3 parameters.
The best fit, shown in Fig.~\ref{fig:pt}(a), gives \PW\ boson yields that are in agreement to within 3\% with the number of events found in the main analysis.

Muon detector acceptance is evaluated with a sample of \PW\ events generated with \PYTHIA and CTEQ6L~\cite{Pumplin:2002vw} PDF, that contains a weighted mixture of proton-proton, neutron-neutron and neutron-proton interactions, representing the nucleon content of Pb nuclei.

Efficiencies for triggering,
reconstructing, and selecting events are estimated using the \PYTHIA{}+\linebreak[2]\HYDJET samples previously discussed.
Such embedded events, after being processed through full CMS trigger emulation and event reconstruction, reveal that track characteristics, such as the number of hits and the $\chi^2$ for the fit of the hits to muon trajectories, have similar distributions in data and in simulation.
The efficiency is evaluated separately for events with positive and negative muons.

The acceptance parameter $\alpha$ is defined as the fraction of \PW\ bosons generated in the total phase space that decay into a muon with $\abs{\eta^\Pgm}< 2.1$ and $\pt^\Pgm > 25\GeVc$. For \PWp\ decays, $\alpha$ is estimated as 63\%, compared to 49\% for \PWm\ events. The difference in the correction factors has its origin
in the different angular distributions of the positive and negative muons from the \PW\ decays, as discussed in the Introduction.
Within this acceptance, the overall trigger, reconstruction, and selection efficiency $\varepsilon$, including the \PTm\ and \mt\ criteria, averages to $(73 \pm 8)\%$, where the variations with the centrality of the event are considered in the uncertainty. To account for possible detector effects, efficiency corrections are applied as a function of muon pseudorapidity, separately for events with positive and negative muons.

The individual components of muon efficiency are also estimated using data by means of a tag-and-probe method, discussed in Refs.~\cite{Chatrchyan:2011ua, CMS:2010xn}, which entails counting \cPZ\ candidates both with and without applying a probe requirement on one of the muons in $\cPZ \rightarrow \Pgmp \Pgmm$ decays, to estimate: (i) the stand-alone muon-reconstruction efficiency, which is probed using charged tracks from the silicon tracker, (ii) the reconstruction efficiency of the silicon tracker, probed using stand-alone muons, and (iii) the trigger efficiency, probed by measuring the trigger response to the second muon in events triggered by a single-muon requirement. The latter is also checked with high-quality reconstructed muons found in minimum-bias events. In all cases, these efficiencies, estimated with data, agree within the statistical uncertainties with those obtained from simulation.

The total systematic uncertainty on the yield of \PW\ bosons is estimated as 7.4\%, and is defined by summing separate contributions in quadrature as follows. The largest uncertainty is associated with tracking efficiency, and corresponds to 4.9\% accuracy obtained for the tag-and-probe determination of efficiency from data.
Similarly, the uncertainty associated with the muon trigger is estimated as 2.3\%. As indicated previously, the 1\% maximum contribution from unsubtracted background sources is taken as a systematic uncertainty. The procedure used to estimate \PTm\ has a 4\% uncertainty, obtained by examining the impact of changing the threshold, from 1 to 10\GeVc, on the $\pt$ of tracks used to calculate the $\pt$ imbalance.
 The contribution from uncertainties on the calibration and resolution of the $\pt$ of muons is 0.2\%. The probability of misidentifying muon charge is negligible (${\approx}10^{-4 }$) and is ignored. The trigger efficiency for minimum-bias events is known to 3\% accuracy.
The measurement of \PW\ production is performed in the region of phase space defined by $|\eta^\mu|<2.1$ and $\pt^\mu>25$\GeVc, but when the results are extrapolated to a larger region of phase space, using the acceptance corrections detailed above, the total systematic uncertainty rises to 8.2\%. This is due to the 3.5\%
precision on the correction for acceptance that arises from uncertainties on the parameters of the assumed model (\PYTHIA) and the choice of PDF (CTEQ6L) used to create the weighted mixture of NN interactions, as well as uncertainties from
higher-order QCD effects,
ignored higher-order electroweak corrections, and the changes introduced in the underlying kinematics (y, $\pt$) of muons with the modifications made in the associated parameters. Table~\ref{tab:system} presents a summary of the different sources contributing to the total systematic uncertainty in the analysis of PbPb data.

\begin{table}[!ht]
\topcaption{Sources of systematic uncertainties in the analyses of PbPb and \Pp\Pp\ data. The last line of the table contains the uncertainty in luminosity for \Pp\Pp\ data, and the analogous uncertainty in the value of $T_\mathrm{AA}$ for MB events in PbPb data, as described in Section~3.}
\label{tab:system}
\begin{center}
\begin{tabular}{l|c|c}
 \hline
     Sources           & PbPb (\%) & \Pp\Pp\ (\%)  \\ \hline \hline
 MB trigger            & 3.0   & --   \\
 Muon trigger          & 2.3   & 2.0   \\
 Tracking efficiency   & 4.9   & --     \\
 $\pt^\Pgm$ calibration & 0.2   & 0.2   \\
 Isolation             & --    & 1.3  \\
 \PTm  & 4.0   & 2.0  \\
 Background            & 1.0   & 1.0  \\ \hline
 Total experimental    & 7.4   & 3.5  \\ \hline
 Acceptance            & 3.5   & 2.8  \\
 Lumi (or equivalent)  & 5.7   & 6.0  \\
\hline
\end{tabular}
\end{center}
\end{table}

The analysis of \Pp\Pp\ events follows a similar procedure to that described for the PbPb data.
 Triggered events require the presence of a muon with a minimal value of \pt, and those containing a muon reconstructed offline with $\pt^\Pgm > 25\GeVc$ and $\abs{\eta^\Pgm}<2.1$ are accepted, based on the same identification and quality criteria as for PbPb data. In addition, muon isolation is implemented by requiring the scalar sum of the transverse energies deposited in the calorimeters and of the transverse momenta of tracks (excluding the muon candidate) around the axis of the muon, within a cone of radius
$\Delta R=\sqrt{(\Delta\phi)^{2} + (\Delta\eta)^{2}} = 0.3 $, to be below  $0.1 \pt^{\Pgm}$.

Unlike the transverse momentum imbalance based on charged tracks, used in PbPb events, the $\PTm$ in \Pp\Pp\ events is calculated using a particle-flow (PF) technique~\cite{CMS_PFT1} that combines tracking and calorimetric information, with the requirement on \PTm of ${>}20\GeV/c$ implemented for the signal region. The resulting \mt distribution for the muon and \PTm system is shown in Fig.~\ref{fig:pt}(c) (open blue squares), where reasonable agreement is observed with the signal reconstructed in PbPb data.  The apparently-better resolution for \Pp\Pp\
relative to PbPb events can be attributed to
 the more powerful PF technique for measuring the \PTm\ variable and, in general, to fewer particles from the underlying event.
After applying the same criteria as used in the PbPb analysis, a total of 301 $\Pgmp$ and 165 $\Pgmm$ events remain in the \Pp\Pp\ data.

As in the PbPb analysis, the acceptance ($\alpha$) and efficiency ($\varepsilon$) for the inclusive \Pp\Pp\ $\rightarrow \PW X \rightarrow \Pgm \Pgngm X$ processes are evaluated using a sample of reconstructed \PW\ events in \Pp\Pp\ collisions at $\sqrt{s}=2.76\TeV$, simulated with \PYTHIA and processed through the CMS detector simulation. The mean values for acceptance are $\alpha = 61\%$ for \PWp\ and $\alpha = 54\%$ for \PWm\ events, and refer to \PW\ bosons produced in the entire phase space for $\PW \rightarrow \Pgm \Pgngm$ decays, with $\abs{\eta^\Pgm}<2.1$ and $\pt^\mu>25\GeVc$. The average efficiency for selection within this acceptance is $\varepsilon$ = 89\%.

The 3\% background estimated in the selected \Pp\Pp\ sample is based on the same methods used in the PbPb analysis, and, as in the treatment of PbPb collisions, the contribution from electroweak background (2.1 \%) is also subtracted from \Pp\Pp\ data.

The systematic uncertainty on the evaluation of the \Pp\Pp\ cross section is estimated to be 7.0\%, and, as before, obtained by summing in quadrature the uncertainty on luminosity (6\%) and on the efficiency-corrected yield (3.5\%), which is affected by sources similar to those discussed in the PbPb analysis: muon identification and trigger efficiencies (2\%), isolation efficiency (1.3\%), calibration and resolution of muon \pt\ (0.2\%), residual background (1\%), and the calculation of \pt\ (2\%). Uncertainties on $\alpha$ correspond to 2.8\%. Table~\ref{tab:system} presents a summary of the different sources contributing to the total systematic uncertainty.

\section{Results}

 Unless stated otherwise, all results reported in this section are evaluated in the restricted region of phase space defined by $|\eta^\mu|<2.1$ and $\pt^\mu>25$\GeVc.

The yield (N$_{\PW}$) of muons from \PW\ decays per MB event in PbPb collisions, per unit of muon pseudorapidity,
is defined as
$N_{\PW}/\Delta\eta = (N_{\Pgm}^\text{sel} - N^\mathrm{B}_\mathrm{ew}) / (\varepsilon N_\mathrm{MB} \Delta \eta)$,
 where $N_{\Pgm}^\text{sel}$ is the number of selected events, $N^\mathrm{B}_\mathrm{ew}$ is the estimated background from other electroweak processes, $N_\mathrm{MB} = 55.7 \times 10^{6} $ is the number of MB events,
corrected for trigger efficiency, and estimated with an accuracy of 3\%,
$\varepsilon$ is the overall efficiency for \PW\ events, and $\Delta \eta = 4.2$ is the pseudorapidity range used in the analysis.
This yields $N_{\PWp}/\Delta\eta = (159 \pm 10 \pm 12) \times 10^{-8}$ and $N_{\PWm}/\Delta\eta = (154 \pm 10 \pm 12) \times 10^{-8}$, for \PWp\ and \PWm\ events, respectively, where the first uncertainty is statistical and the second is systematic.
Extrapolating these measurements to the total \PW\ phase space using the acceptance corrections reported above, provides the total yields per MB event of $N({\PW})$ = $(1057 \pm 63 \pm 88) \times 10^{-8}$ for \PWp, and $(1317 \pm 80 \pm 108) \times 10^{-8}$ for \PWm.

The above analysis is repeated after subdividing the data into the five bins of event centrality (defined in Section~2) and six bins in muon $|\eta|$. The total systematic uncertainty does not depend significantly on these variables, and is considered to be constant and uncorrelated between bins.

The yields of muons from \PW\ decays per MB event and per unit of muon pseudorapidity, $N_{\PW}/\Delta\eta$, in PbPb collisions can be turned  into the inclusive \PW\ cross sections per unit of muon pseudorapidity,
normalised to the number of binary collisions occurring in PbPb interactions, when divided by the scaling factor $T_\mathrm{AA}$, that is,
$(1/T_\mathrm{AA}) (N_\PW/\Delta\eta)$. This factor represents the nuclear-overlap function, namely the number of elementary binary NN collisions divided by the elementary NN cross section, and can be interpreted as the NN-equivalent integrated luminosity per AA collision for a given centrality.
In units of collisions per mb, the average $T_\mathrm{AA}$ corresponds to $0.47 \pm 0.07$, $3.9 \pm 0.4$, $8.8 \pm 0.6$, $14.5 \pm 0.8$, and $23.2 \pm 1.0$, for event centralities of 50--100\%, 30--50\%, 20--30\%, 10--20\% and 0--10\%, respectively, and to $5.66 \pm 0.32$ for MB events, as
computed using a Glauber model~\cite{Miller:2007ri}, with the parameters given in Ref.~\cite{HIN-10-004}. The quoted uncertainties are obtained by changing the parameters of the model and the MB trigger and selection efficiencies by their respective uncertainties.

\begin{figure}[hbtp]
  \begin{center}
    \includegraphics[width=\cmsFigWidth]{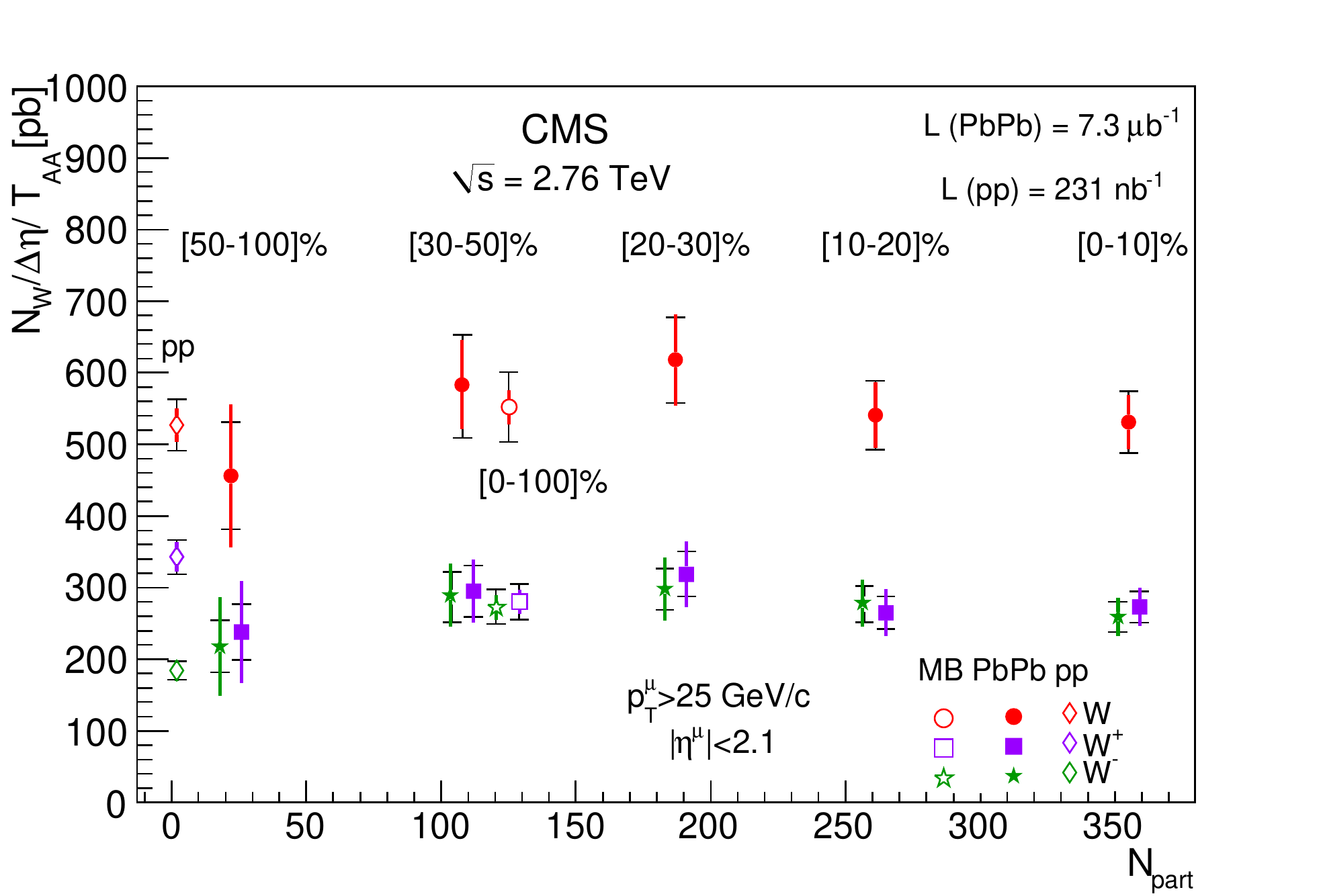} % 0.95 for PRL
    \caption{Centrality dependence of normalised \PW $\rightarrow \mu \Pgngm$ cross sections $(1/T_\mathrm{AA}) (N_\PW/\Delta\eta)$ in PbPb collisions,
for all \PW\ candidates (red-filled points) and,
separated by charge, \PWp\ (violet-filled squares) and \PWm\ (green-filled stars).
The open symbols at $N_\text{part} \approx 120 $ represent the MB events.  At $N_\text{part} = 2$, the corresponding cross sections are displayed for \Pp\Pp\ collisions divided by $\Delta \eta$, for the same $\sqrt{s}$. For clarity, both \PWp\ and \PWm\ points are slightly shifted on the horizontal axis. The cross sections are given for the phase space region $|\eta^\Pgm|<2.1$ and $\pt^\Pgm > 25$\GeVc. The error bars represent the statistical, and the horizontal lines the systematic, uncertainties.}
    \label{fig:cent}
  \end{center}
\end{figure}

Figure~\ref{fig:cent} shows the centrality dependence of the separate \PWp\ (violet-filled squares) and \PWm\ (green-filled stars) normalised (NN-equivalent)
cross sections and their sum (red-filled circles).
The abscissa represents the average number of participating nucleons that undergo inelastic hadronic interactions ($N_\text{part}$) for the selected centrality intervals and computed using the same Glauber model. The open symbols at $N_\text{part} \approx 120$ correspond to MB events. For clarity, both \PWp\ and \PWm\ points are slightly shifted relative to each other on the horizontal axis. Within present uncertainties, the normalised cross sections of \PW\ bosons in terms of the number of elementary nucleon-nucleon collisions are consistent with being independent of centrality of the PbPb collision, as can be seen by the results shown in Fig.\ref{fig:cent} and Table~\ref{tab:Yields_centr}.

\begin{table*}[!htb]
\begin{center}
\topcaption{ The selected number of events with $\Pgmp$ ($N_{\Pgmp}^\text{sel}$) and $\Pgmm$ ($N_{\Pgmm}^\text{sel}$) and the normalised cross sections $(1/T_\mathrm{AA}) (N_\PW/\Delta\eta)$ as a function of centrality in PbPb data. The last row provides the selected number of events with $\Pgmp$ ($N_{\Pgmp}^\text{sel-\Pp\Pp}$) and $\Pgmm$ ($N_{\Pgmm}^\text{sel-\Pp\Pp}$) in \Pp\Pp\ data, and the cross sections derived from them (see text), divided by $\Delta\eta$. All values are reported for $|\eta^\Pgm|<2.1$ and $\pt^\Pgm> 25\GeVc$.}
\label{tab:Yields_centr}
\begin{tabular}{c|c|c|c|c}
 \hline
Centrality (\%) & $N_{\Pgmp}^\text{sel}$ & Normalised cross section ($\PWp$) & $N_{\Pgm}^\text{sel}$ & Normalised cross section ($\PWm$)  \\
 &   &  [nb]/$\Delta\eta$ & & [nb]/$\Delta\eta$ \\ \hline \hline
(50--100) & 11  & 0.24 $\pm$ 0.07 $\pm$ 0.04 & 10  & 0.22 $\pm$ 0.07 $\pm$ 0.04 \\
(30--50)  & 45  & 0.30 $\pm$ 0.04 $\pm$ 0.04 & 43  & 0.28 $\pm$ 0.04 $\pm$ 0.04 \\
(20--30)  & 48  & 0.32 $\pm$ 0.05 $\pm$ 0.03 & 46  & 0.30 $\pm$ 0.04 $\pm$ 0.03 \\
(10--20)  & 66  & 0.27 $\pm$ 0.03 $\pm$ 0.02 & 72  & 0.28 $\pm$ 0.03 $\pm$ 0.02 \\
(0--10)   & 105 & 0.27 $\pm$ 0.03 $\pm$ 0.02 & 93  & 0.26 $\pm$ 0.03 $\pm$ 0.02 \\ \hline
(0--100)  & 275 & 0.28 $\pm$ 0.02 $\pm$ 0.02 & 264 & 0.27 $\pm$ 0.02 $\pm$ 0.02 \\ \hline
 \Pp\Pp\  & 301 & 0.34 $\pm$ 0.02 $\pm$ 0.02 & 165 & 0.18 $\pm$ 0.01 $\pm$ 0.01 \\ \hline
\end{tabular}
\end{center}
\end{table*}

The \PW-production cross sections for \Pp\Pp\ collisions at the same $\sqrt{s}$ are determined in a similar manner, according to $ \sigma_{\Pp\Pp} = \sigma (\Pp\Pp\rightarrow \PW X ) \cdot B(\PW \rightarrow \Pgm \Pgngm) = (N_{\Pgm}^\text{sel-pp} - N^\mathrm{B}_\mathrm{ew}) / (\varepsilon_{\Pp\Pp} \mathcal{L})$, where $B$ is the \PW\ boson leptonic decay branching fraction, $N_{\Pgm}^\text{sel-\Pp\Pp}$ is the number of selected events from the \Pp\Pp\ data sample, $\varepsilon_{\Pp\Pp}$ the overall efficiency for \PW\ bosons, and $\mathcal{L}$ the total integrated \Pp\Pp\ luminosity of 231\nbinv. The corresponding values, divided by the muon pseudorapidity interval of $\Delta \eta = 4.2$, are displayed at $N_\text{part} = 2$ in Fig.~\ref{fig:cent} as open diamonds, and with the same colour code as for the PbPb data. The numerical values are given in Table~\ref{tab:Yields_centr}.
The impact of the neutron content in Pb nuclei is observed in
the enhancement of \PWm\ and the reduction of \PWp\ production in PbPb relative to \Pp\Pp\ interactions.
As can be seen,
the individual \PWp\ and \PWm\ cross sections differ,
while their sum agrees for PbPb and \Pp\Pp\ collisions.

Theoretical predictions for \PW-boson production in PbPb collisions are based on the next-to-leading-order (NLO) \MCFM~\cite{MCFM} calculation and MSTW2008 PDF~\cite{PDF4LHC} at next-to-NLO (NNLO), interfaced with the EPS09~\cite{Paukkunen:2010qg} nuclear PDF that account for nuclear modifications in collisions involving heavy ions rather than a sum over free nucleons. The effect of this nuclear PDF package is a 4\% reduction on the total \PW\ cross section with respect to the free nucleon PDF.
The cross sections per NN pair, obtained in the region of phase space studied in this analysis, are
$(0.97 \pm 0.10)\unit{nb}$ for \PWp\ and $(0.87 \pm 0.09)\unit{nb}$ for \PWm\ production,
where the quoted uncertainties take account of the two choices of PDF~\cite{Paukkunen:2010qg}.
Dividing these values by $\Delta\eta$ = 4.2, provides predictions of $(0.23 \pm 0.02)\unit{nb}$ for \PWp\ and $(0.21 \pm 0.02)\unit{nb}$ for \PWm\ production per unit of muon pseudorapidity, values that are compatible with the experimental results given in Table~\ref{tab:Yields_centr}.

The cross sections for pp$\rightarrow \PW X \rightarrow \Pgm \Pgngm X$ given above are now presented in the region of muon acceptance, denoted by $\sigma_\text{acc}\cdot B$, and are as follows:

\ifthenelse{\boolean{cms@external}}
{
\begin{align*}
 &\sigma_\text{acc}(\Pp\Pp\rightarrow \PW^{+}X)\cdot B(\PWp \rightarrow \Pgmp \Pgngm)  = \\
 &\quad (1.44 \, \pm \, 0.08 \, \pm \, 0.10)\unit{nb};\\
 &\sigma_\text{acc}(\Pp\Pp\rightarrow \mathrm{W}^-X)\cdot B(\PWm \rightarrow \Pgmm \Pagngm ) = \\
 &\quad (0.77 \, \pm \, 0.06 \, \pm \, 0.05)\unit{nb}; \\
 &\sigma_\text{acc}(\Pp\Pp\rightarrow \mathrm{W} X)\cdot B(\PW \rightarrow \mu \Pgngm) = \\
 &\quad (2.22 \, \pm \, 0.10 \, \pm \, 0.16)\unit{nb}.
\end{align*}
}{
\begin{equation*}
\begin{array}{rcl}
 \sigma_\text{acc}(\Pp\Pp\rightarrow \PW^{+}X)\cdot B(\PWp \rightarrow \Pgmp \Pgngm)  & = & (1.44 \, \pm \, 0.08 \, \pm \, 0.10)\unit{nb};\\
\sigma_\text{acc}(\Pp\Pp\rightarrow \mathrm{W}^-X)\cdot B(\PWm \rightarrow \Pgmm \Pagngm ) & = & (0.77 \, \pm \, 0.06 \, \pm \, 0.05)\unit{nb}; \\
\sigma_\text{acc}(\Pp\Pp\rightarrow \mathrm{W} X)\cdot B(\PW \rightarrow \mu \Pgngm) & = & (2.22 \, \pm \, 0.10 \, \pm \, 0.16)\unit{nb}.
\end{array}
\end{equation*}
}

Predictions for \PW-boson production in \Pp\Pp\ collisions at $\sqrt{s} = 2.76\GeV$ are obtained using the NNLO program \textsc{fewz}~\cite{FEWZ} and the NNLO MSTW2008 PDF, and correspond to cross sections of (1.32 $\pm$ 0.06) nb for \PWp\ and $(0.72 \pm 0.03)\unit{nb}$ for \PWm\ in the same region of study.
The uncertainties are from the choice of PDF.

The experimental results extrapolated to the entire phase space of \PW-boson production, using the previous acceptance corrections, and referred to as $\sigma_\text{tot}\cdot B$, are:

\ifthenelse{\boolean{cms@external}}
{
\begin{align*}
&\sigma_\text{tot}(\Pp\Pp\rightarrow \PWp X)\cdot B(\PWp \rightarrow \Pgmp \Pgngm)   = \\
&\quad (2.38 \, \pm \, 0.14 \, \pm \, 0.18)\unit{nb}; \\
&\sigma_\text{tot}(\Pp\Pp\rightarrow \PWm X)\cdot B(\PWm \rightarrow \Pgmm \Pagngm) = \\
&\quad (1.45 \, \pm \, 0.11 \, \pm \, 0.11)\unit{nb}; \\
&\sigma_\text{tot}(\Pp\Pp\rightarrow \PW X)\cdot B(\PW \rightarrow \mu \Pgngm) = \\
&\quad(3.83 \, \pm \, 0.18 \, \pm \, 0.29)\unit{nb}.
\end{align*}
}{
\begin{equation*}
\begin{array}{rcl}
\sigma_\text{tot}(\Pp\Pp\rightarrow \PWp X)\cdot B(\PWp \rightarrow \Pgmp \Pgngm)  & = & (2.38 \, \pm \, 0.14 \, \pm \, 0.18)\unit{nb}; \\
\sigma_\text{tot}(\Pp\Pp\rightarrow \PWm X)\cdot B(\PWm \rightarrow \Pgmm \Pagngm) & = & (1.45 \, \pm \, 0.11 \, \pm \, 0.11)\unit{nb}; \\
\sigma_\text{tot}(\Pp\Pp\rightarrow \PW X)\cdot B(\PW \rightarrow \mu \Pgngm) & = & (3.83 \, \pm \, 0.18 \, \pm \, 0.29)\unit{nb}.
\end{array}
\end{equation*}
}

The corresponding NNLO predictions from \textsc{fewz} are $(2.11 \pm 0.10)\unit{nb}$ for \PWp\ and $(1.29 \pm 0.05)\unit{nb}$ for \PWm\ production.
The experimental results are compatible with the predictions, thereby confirming the validity of the standard model for \PW\ production at $\sqrt{s} = 2.76$\TeV.

The nuclear modification factors $R_\mathrm{AA}=N_\PW / (T_\mathrm{AA} \cdot \sigma_{\Pp\Pp})$, relating the \PW\ production in PbPb and in \Pp\Pp\ collisions, are computed from the measured yields in PbPb ($N_\PW$), the $\Pp\Pp\rightarrow \PW X \rightarrow \Pgm \Pgngm X$ measured cross sections, both quantities in the region of muon acceptance, and the nuclear-overlap function ($T_\mathrm{AA}$).
Although the overall \PW-production cross section is found to scale with the number of elementary collisions, the individual \PWp\ and \PWm\ yields show a strong modification due to the nucleon content in Pb nuclei, as indicated by the extracted $R_\mathrm{AA}$ factors for the region of phase space studied:

\begin{equation*}
\begin{array}{rcl}
R_\mathrm{AA}(\PWp) & = &  0.82 \, \pm \, 0.07 \, \pm \, 0.09; \\
R_\mathrm{AA}(\PWm) & = &  1.46 \, \pm \, 0.14 \, \pm \, 0.16; \\
R_\mathrm{AA}(\PW)   & = &  1.04 \, \pm \, 0.07 \, \pm \, 0.12.
\end{array}
\end{equation*}

No cancellation of systematic uncertainties is assumed in computing these ratios.

\begin{figure}[hbtp]
  \begin{center}
    \includegraphics[width=\cmsFigWidth]{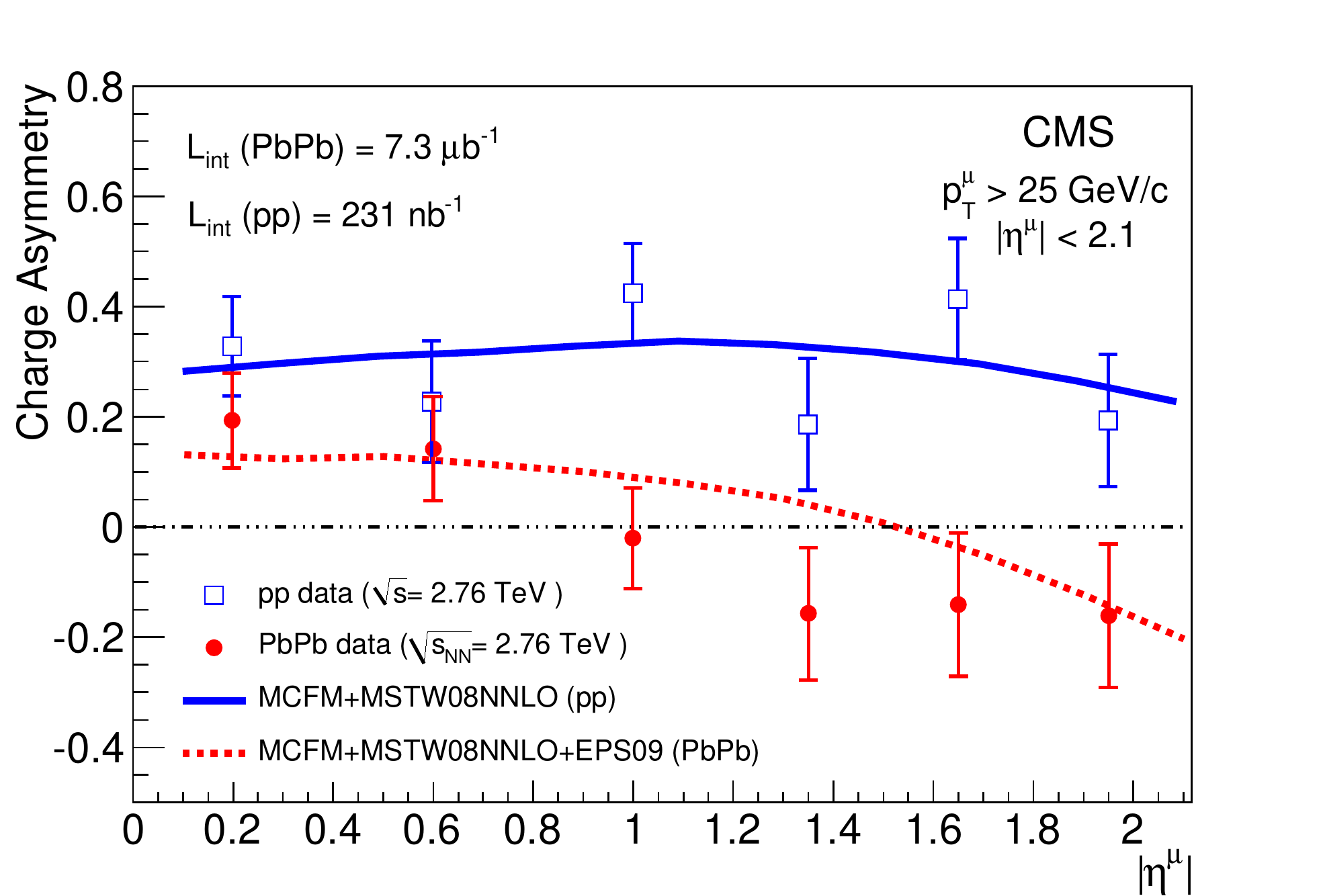} % 0.95 for PRL
    \caption{Charge asymmetry $(N_{\PWp} - N_{\PWm}) / (N_{\PWp} + N_{\PWm})$ as a function of muon pseudorapidity for PbPb (red-filled circles) and \Pp\Pp\ (blue open squares) collisions at $\sqrt{s} = 2.76\TeV$. Overlaid are predictions for \Pp\Pp\ collisions from \MCFM calculations with MSTW2008 PDF (blue solid curve), as well as expectations for PbPb collisions from \MCFM with MSTW2008 and EPS09 PDF, that include isospin and nuclear effects (red dashed curve). Each prediction has an additional uncertainty of 5\%, estimated by the uncertainty from the choice of PDF. The experimental points have an additional 1.1\% systematic uncertainty that is not shown in the figure. The dot-dashed horizontal line is drawn only for reference. }
    \label{fig:asym}
  \end{center}
\end{figure}

The difference in \PWp\ and \PWm\ production at LHC and their subsequent leptonic decays provide a
different yield of \PW$^+ \rightarrow \Pgmp \Pgngm$ and \PW$^- \rightarrow \Pgmm \Pagngm$,
defined as the "charge asymmetry" and given by
$A = (N_{\PWp} - N_{\PWm}) / (N_{\PWp} + N_{\PWm})$, where N$_{\PW}$ represents the efficiency-corrected  number of selected events with a muon from a \PW\ decay, once the background from other electroweak processes has been subtracted.
Figure~\ref{fig:asym} shows this difference
as a function of the muon pseudorapidity for PbPb collisions at 2.76\TeV (red-filled circles), for the experimental region studied. The dot-dashed horizontal line at zero asymmetry is drawn for reference.
Although the uncertainties are quite large, the measured asymmetry changes from positive to negative values, which indicates an excess of \PWm\ over \PWp\ production at large $|\eta^\mu|$.
The dependence of the asymmetry on muon pseudorapidity is in agreement with the predictions from \MCFM and the MSTW2008 PDF, with small additional nuclear effects provided through the EPS09 nuclear PDF~\cite{Paukkunen:2010qg}, and represented by the dashed curve. The uncertainty on the prediction associated to the use of both PDF ranges from ${\approx}3\%$ to ${\approx}8\%$ between small and large muon pseudorapidities.

Most of the systematic uncertainties on these measurements affect \Pgmp\ and \Pgmm\ events equally, and tend to cancel in the ratio $A$. A residual effect remains from a statistics-limited difference in efficiency observed for positive and negative muons in certain regions of pseudorapidity. The maximum difference (0.5\%) is assigned as a systematic uncertainty. The effect of a possible difference in the calibration of muon $\pt$ or in the resolution for oppositely charged leptons is evaluated to be below $0.2\%$, and the impact of the subtraction of backgrounds from QCD processes is estimated to be of 1\%. These factors amount to a common systematic uncertainty of 1.1\% that affects each measured point.

 Results for pp collisions at 2.76\TeV
are also shown in Fig.~\ref{fig:asym} (blue open squares), together with predictions from \MCFM using the MSTW2008 PDF, represented by the solid curve. The uncertainty on the theoretical prediction is ${\approx} 5\%$, which reflects the uncertainty from the choice of PDF. Central values obtained with \MCFM, but using other PDF (CT10, CTEQ6.6M)~\cite{CTEQ_PDF}, differ by ${\approx}10\%$ from the result shown in Fig.~\ref{fig:asym}.

Effects due to the nucleon content in Pb nuclei are clearly visible in comparing results from PbPb and \Pp\Pp\ collisions. For the latter, the yield of $\Pgmp$ exceeds that of $\Pgmm$ at all pseudorapidities, reflecting the dominance of \PWp\ over \PWm\ production.
 The integrated charge asymmetry $A = 0.30 \pm 0.04 \pm 0.01$ is in agreement with the \MCFM prediction using MSTW2008 PDF, which yields $A = 0.30 \pm 0.03$.
Recent measurements of this quantity in
pp collisions at $\sqrt{s} =7\TeV$~\cite{Chatrchyan:2011jz}, for the fiducial region analysed in this Letter, yield $A = 0.189 \pm 0.002 \pm 0.008$,
  indicating a dependence of the asymmetry on the CM energy of the interaction, which is in agreement with expectations from \PW\ kinematics and PDF evolution~\cite{PDF4LHC}.

\section{Conclusions}

The inclusive production of \PW\ bosons has been measured for $\PW \rightarrow \mu \Pgngm$ decays in PbPb and \Pp\Pp\ collisions at $\sNN = 2.76$\TeV. Decays of \PW\ bosons
 were identified by requiring a reconstructed muon with $|\eta^\mu|<2.1$
and $\pt^{\mu}>25\GeVc$ and transverse mass $\mt > 40\GeVcc$.
The \PW\ yields for all PbPb collision centralities were found to be consistent
with those measured in \Pp\Pp\ collisions scaled by the corresponding number of
incoherent nucleon-nucleon interactions. The individual \PWp\ and \PWm\ boson yields
are modified in PbPb compared to \Pp\Pp\ collisions due to the different proton and
neutron content in the nuclear beams.  The differences in the $\Pgmp$ and $\Pgmm$
yields from \PW\ decays (charge asymmetries) have been measured as a function of
muon pseudorapidity both in PbPb and \Pp\Pp\ collisions.
All measurements were found to be well reproduced by
higher-order perturbative QCD predictions.
The \Pp\Pp\ results combined with those obtained at $\sqrt{s} = 7\TeV$, exhibit the
expected dependence of the charge asymmetry on the parton densities as probed
at different collision energies.

The results confirm the theoretical expectation that, in the probed ranges of
parton fractional momentum and energy scale, further modifications of the
nuclear parton distribution functions in the lead nucleus compared to the
proton are small relative to the dominant isospin effect.

These studies demonstrate the promise of \PW-boson measurements as powerful
tools in the investigation of initial and final-state effects in nuclear collisions at
the LHC. The charge asymmetry of \PW-boson yields
in PbPb interactions provides unique sensitivity to the parton distribution
functions for neutrons.
Future analyses of larger data samples will yield
enhanced constraints on the parton densities in nuclei and allow studies
of \PW\ production in association with jets.

\section*{Acknowledgments}

We thank Carlos Salgado for fruitful theoretical inputs on nuclear effects in \PW\ production.
We congratulate our CERN accelerator-department colleagues for the excellent performance of the LHC machine in 2010 and thank the technical and administrative staffs at CERN and other CMS institutes for their contributions, and acknowledge support from: FMSR (Austria); FNRS and FWO (Belgium); CNPq, CAPES, FAPERJ, and FAPESP (Brazil); MES (Bulgaria); CERN; CAS, MoST, and NSFC (China); COLCIENCIAS (Colombia); MSES (Croatia); RPF (Cyprus); Academy of Sciences and NICPB (Estonia); Academy of Finland, ME, and HIP (Finland); CEA and CNRS/IN2P3 (France); BMBF, DFG, and HGF (Germany); GSRT (Greece); OTKA and NKTH (Hungary); DAE and DST (India); IPM (Iran); SFI (Ireland); INFN (Italy); NRF and WCU (Korea); LAS (Lithuania); CINVESTAV, CONACYT, SEP, and UASLP-FAI (Mexico); PAEC (Pakistan); SCSR (Poland); FCT (Portugal); JINR (Armenia, Belarus, Georgia, Ukraine, Uzbekistan); MST and MAE (Russia); MSTD (Serbia); MICINN and CPAN (Spain); Swiss Funding Agencies (Switzerland); NSC (Taipei); TUBITAK and TAEK (Turkey); STFC (United Kingdom); DOE and NSF (USA). Individuals have received support from the Marie-Curie programme and the European Research Council (European Union); the Leventis Foundation; the A. P. Sloan Foundation; the Alexander von Humboldt Foundation; the Belgian Federal Science Policy Office; the Fonds pour la Formation \`a la Recherche dans l'Industrie et dans l'Agriculture (FRIA-Belgium); the Agentschap voor Innovatie door Wetenschap en Technologie (IWT-Belgium); and the Council of Science and Industrial Research, India.

\bibliography{auto_generated}   % will be created by the tdr script.

\cleardoublepage \appendix\section{The CMS Collaboration \label{app:collab}}\begin{sloppypar}\hyphenpenalty=5000\widowpenalty=500\clubpenalty=5000\textbf{Yerevan Physics Institute,  Yerevan,  Armenia}\\*[0pt]
S.~Chatrchyan, V.~Khachatryan, A.M.~Sirunyan, A.~Tumasyan
\vskip\cmsinstskip
\textbf{Institut f\"{u}r Hochenergiephysik der OeAW,  Wien,  Austria}\\*[0pt]
W.~Adam, T.~Bergauer, M.~Dragicevic, J.~Er\"{o}, C.~Fabjan\cmsAuthorMark{1}, M.~Friedl, R.~Fr\"{u}hwirth\cmsAuthorMark{1}, V.M.~Ghete, J.~Hammer, N.~H\"{o}rmann, J.~Hrubec, M.~Jeitler\cmsAuthorMark{1}, W.~Kiesenhofer, V.~Kn\"{u}nz, M.~Krammer\cmsAuthorMark{1}, D.~Liko, I.~Mikulec, M.~Pernicka$^{\textrm{\dag}}$, B.~Rahbaran, C.~Rohringer, H.~Rohringer, R.~Sch\"{o}fbeck, J.~Strauss, A.~Taurok, P.~Wagner, W.~Waltenberger, G.~Walzel, E.~Widl, C.-E.~Wulz\cmsAuthorMark{1}
% ...
\vskip\cmsinstskip
\dag:~Deceased\\
1:~~Also at Vienna University of Technology, Vienna, Austria\\
% ...

\end{sloppypar}
\end{document}